# A new method to suppress high-order harmonics for synchrotron radiation soft x-ray beamline*

GUO Zhi-Ying(郭志英)[1] HONG Cai-Hao(洪才浩)[1] XING Hai-Ying(邢海英)[2] TANG Kun(唐坤)[1] HAN Yong(韩勇)[1]

CHEN Dong-liang(陈栋梁)[1] ZHAO Yi-Dong(赵屹东)[1)]

1 Institute of High Energy Physics, Chinese Academy of Sciences, Beijing 100049, China

2 School of Electronics and Information Engineering, Tianjin Polytechnic University, Tianjin 300387, China

**Abstract:** A feasible and convenient method has been proposed to suppress higher-harmonics for varied-line-spacing (VLS) plane grating monochromator in soft x-ray region. Related calculations and experiments demonstrate that decreasing the included angle slightly by changing the parameter of exit arm length can significantly improve light purity. This method is suitable and has been used for experiments of detector calibration in beamline 4B7B at Beijing Synchrotron Radiation Facility (BSRF).

**Keywords:** soft x-ray; high-order harmonics; detector calibration; Monk-Gillieson monochromator; synchrotron radiation;

**PACS:** 41.50.+h, 06.20.fb, 41.85.Si, 07.85.Qe

## 1　Introduction

The double crystal monochromator (DCM) and grating monochromator have been widely used in synchrotron radiation beamline, but both of them have a problem of high-order harmonics in the monochromatic lights. For hard x-ray, the problem can be overcome by slightly detuning the second crystal of the DCM [1]. However, for soft x-rays it is difficult to suppress harmonics. So far, many efforts have been used to solve the problem. These can be classified into three kinds of methods based on their techniques: First, select appropriate materials to use as transmission filters [2] [3], it works because the ratio of each harmonic in the transmitted light are changed by the absorption edge of a particular element; Second, use several pairs of total reflection mirrors as suppressor [4] [5], this method can cover a wider energy range, but it is more difficult to align the optical mirrors compared to the first method, and the light path may changes slightly when switch to another set of mirrors; Third, change the modes of operation or replaced the grating with a designated one. This method is often used in the SX-700 [6] [7] monochromator and varied-line-spacing plane grating monochromator (VSPG) in soft x-ray range. SX-700-II [8] can operate in higher order suppression (HIOS) mode by change the parameter of $C_{ff}$[9]. VSPG monochromator can change another grating in order to get the high performance of spectral purity. However, these methods are expensive and not convenient enough, so we proposed an easy method which just likes the DCM detuning method to suppress the harmonics for VSPG monochromator. Decreasing the included angle slightly by changing the parameter of exit arm length can significantly improve light purity. This method has been used in beamline 4B7B at BSRF for detector calibration experiments.

* Supported by National Natural Science Foundation of China (11375227, 61204008)
1) E-mail: zhaoyd@ihep.ac.cn

4B7B is a soft x-ray beamline for detector calibration experiment and light element absorption spectroscopy, which used a variable-included-angle Monk-Gillieson mounting monochromator with a varied-line-spacing plane grating covering the energy range of 50eV~1600eV. The content of higher orders light varies with the X-ray energy. When the energy falls within 50-500eV, the problem of harmonics is serious especially between 50-150eV. For detector calibration experiments the light purity is important so it is necessary to develop a new method to reduce the harmonics. With energy lager than 800eV, it is pure enough because the optical mirrors offer high reflectivity for low energy X-rays. So the harmonic problem should be especially considered at the low energy range. Our novel method works well and related calculations are listed at the second section. The third section described the relevant experiment which was used to verify our method. The discussion and conclusion are listed at the end of the article.

## 2  Analysis and calculation method

The content of each harmonic order and the corresponding suppression ratio are analyzed and calculated by the XOP software. These calculations are based on the optical components and the structures of 4B7B beamline which are shown in Fig 1.

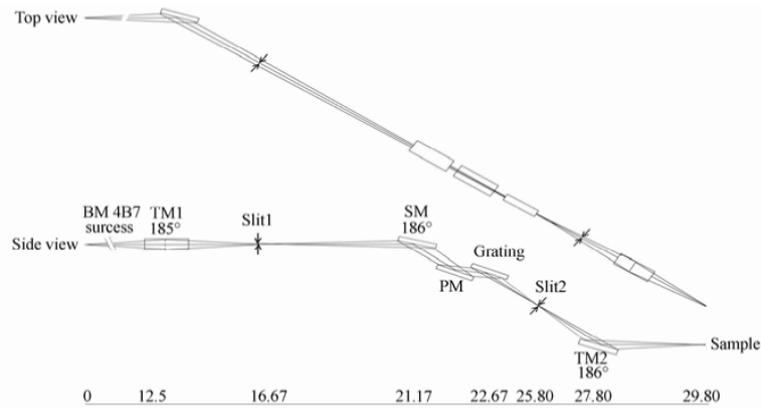

Fig.1 The layout of the 4B7B beamline

The flux of first, second and third harmonic at the sample can be given as follows:

$$Flux_{sample}(\lambda/n) = \varepsilon \times Flux_{source}(\lambda/n) \qquad (1)$$
$$\varepsilon = c \cdot \varepsilon_{TM1}(\theta_{TM1},\lambda/n) \times \varepsilon_{SM}(\theta_{SM},\lambda/n) \times \varepsilon_{TM2}(\theta_{TM2},\lambda) \times \varepsilon_{G}(\alpha,\beta,\lambda) \times \varepsilon_{PM}(\theta_{PM},\lambda)$$

where $\varepsilon$ is the total transmission efficiency of the beamline. $\varepsilon_{TM1}$, $\varepsilon_{PM}$, $\varepsilon_{TM2}$ are the reflectivity of the first toroid mirror, plane mirror and the second toroid mirror. $\varepsilon_{G}$ is the diffraction efficiency of the laminar grating, and it depends on the incident angle

$\alpha$, the diffraction angle $\beta$ and the wavelength of the x-ray if other parameters of the grating have been fixed. $\theta_{TM1}$, $\theta_{TM2}$, $\theta_{SM}$ are the incident angles of the first, second toroid mirror and the sphere mirror. The incident and diffraction angles of the grating depend on the working curve and vary with wavelength. $\theta_{PM}$ is the incidence angle of the plane mirror. $\lambda/n$ is the wavelength of each harmonic order (first order n=1, second order n=2…).Considering the acceptance efficiency of the beamline and the transmission efficiency of the slits, a constant factor *c* are introduced. The reflectivity of the mirror can be calculated with XOP 2.1[10].The results show that the reflectivity of second order light decrease rapidly when the energy is above 500eV, while third order is 800eV. The diffraction efficiency of the grating was calculated using the differential formalism of the exact electromagnetic theory[11].The results are illustrated in the Fig 2, which demonstrates the efficiency of second order is higher in the range of 70eV-120eV and the third order is more serious in the long wavelength. According to the above formula, we got the suppression ratio with different energy in Fig 3.

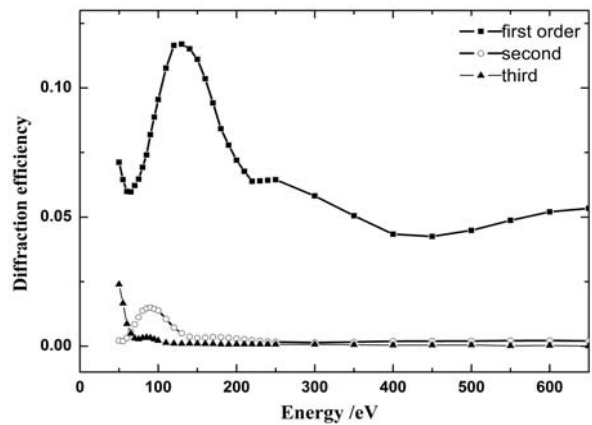

Fig. 2 The diffraction efficiency of the laminar grating of 800*l/mm*

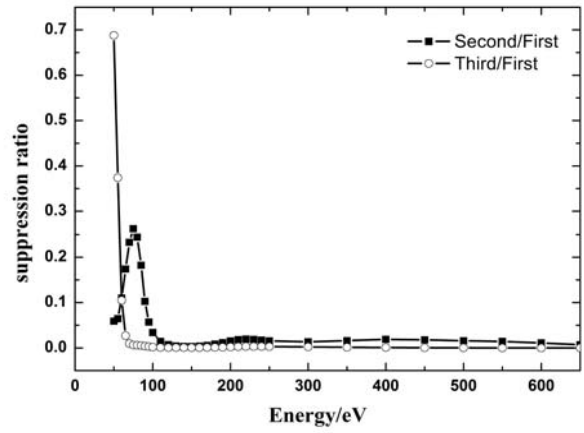

Fig.3 The suppression ratio with different energy

The grating equation and meridional focus equation [12] of this kind of beamline are:

$$d_0(\sin\alpha + \sin\beta) = m\lambda$$
$$\left(\frac{\cos^2\alpha}{r} + \frac{\cos^2\beta}{r'}\right) + b_2\frac{m\lambda}{d_0} = 0 \qquad (2)$$

Adjusting the exit arm parameter $r'$ smaller, we find the harmonic suppression ratio smaller. In other words, the spectral purity improves. This phenomenon can be explained by the decrease of the included angle $\alpha - \beta$ based on the above equation. The reflectivity of first order declined even faster than harmonics. In order to clearly demonstrate this change, we calculate how the suppression ratio change with the $r'$ at 170eV. The result is shown in Fig 4. The suppression ratio will decrease from 0.54% to 0.25% if the parameter changes from 3165*mm* to 3145*mm*.

At different energy the curve is different more or less, but the trend is the same. So appropriate reduction of the value $r'$ is benefit for harmonics suppress ratio. However the resolution and flux will be reduced accordingly. Since other methods sacrificed flux also and the resolution of the beamline is high enough for detector calibration experiments, we still believe it is a good method to suppress harmonics.

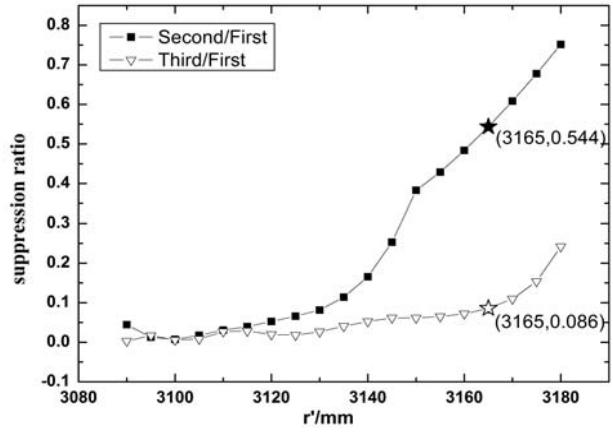

Fig 4. Harmonics suppress ratio with parameter $r'$

## 3  Experiments

The effect of this method was confirmed by the total electron yield (TEY) spectra. Similar experiment [9] has been done for SX-700 to verify the effect of HIOS. The absorb spectroscopy of $CaSO_4$ was collected at TEY mode on 4B7B beamline at BSRF. The electron energy was 2.5Gev and the beam current was 150-250mA. The resolution power ($E/\Delta E$) of the beamline is better than 7000 at 90eV and 6000 at 250eV. During date collection, the spectra signal is normalized by the TEY signal measured from a copper foil located under the second toroid mirror. The sample was measured at different values of $r'$. The linear extrapolation of the pre-edge has been subtracted as background. The spectra have been normalized to the absorption peak of sulfur. As shown in Fig 5, the $L_{23}$ edge of calcium appeared in the spectra because of the second order light and it showed significant changes with the parameter $r'$. The sharp peak of Ca got more and more weaker and smaller, which demonstrate the content of high orders light is suppressed .This experiment result is consistent with the theoretical result in Fig4.

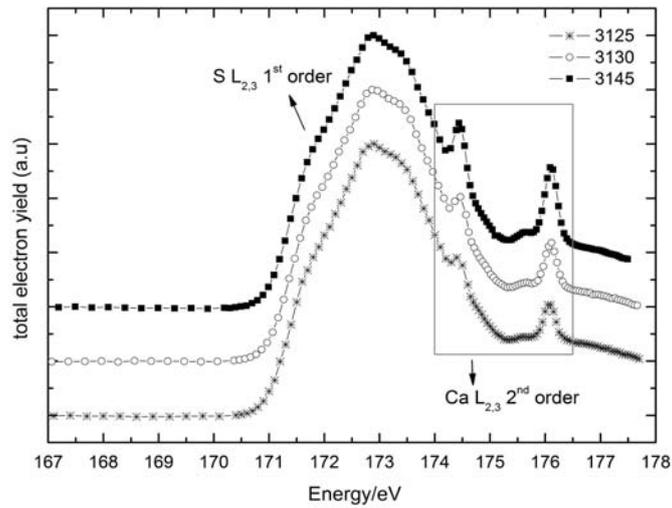

Fig5. Total electron yield spectra of $CaSO_4$ for the three different values of $r'$, and it shows that the optical purity is improved by decreasing the parameter $r'$.

## 4 Discussion and conclusion

The experimental and calculation results show that the introduced approach improves the light purity. And it shows feasible and convenient and can be combined with other harmonic suppression method in calibration experiments. The most appropriate parameter can be found for different photon energy if a compromise between the resolution and purity are taken into account. In addition, a quantitative measurement of the proportions of high-order harmonics [13] can make a further verification.